\documentclass[%
 notitlepage,
 twocolumn,
superscriptaddress,
 amsmath,
 amssymb,
 aps,
 pra,
]{revtex4-1}
\usepackage{amsfonts,amssymb,amsmath,physics,graphicx}
\usepackage[T1]{fontenc}
\usepackage[utf8]{inputenc}
\usepackage[english]{babel}
\usepackage{soul}
\usepackage{soul}

\usepackage{hyperref}
\hypersetup{
citecolor=blue,
 colorlinks=true,
 urlcolor=magenta,
}

\renewcommand{\bra}[1]    {\langle #1|}
\renewcommand{\ket}[1]    {|#1 \rangle}
\newcommand{\ketL}[1]    {\left|#1 \right\rangle}
\renewcommand{\ketbra}[2]{|#1\rangle\!\langle#2|}

\newcommand{\av}[1]    {\langle #1 \rangle}

\newcommand{\modsq}[1]    {\left| #1 \right|^2}

\newcommand{\ICFO}{ICFO - Institut de Ciencies Fotoniques, The Barcelona Institute of Science and Technology, 08860 Castelldefels, Barcelona, Spain}
\newcommand{\ICREA}{ICREA, Pg. Lluís Companys 23, 08010 Barcelona, Spain}
\newcommand{\IOPPAS}{Institute of Physics PAS, Aleja Lotnikow 32/46, 02-668 Warszawa, Poland}
\newcommand{\UW}{Faculty of Physics, University of Warsaw, ul. Pasteura 5, PL-02-093 Warsaw, Poland}

\begin{document}

\title{One-axis twisting as a method of generating many-body Bell correlations}

\author{Marcin P\l{}odzie\'n}
\affiliation{\ICFO}

\author{Maciej Lewenstein}
\affiliation{\ICFO}
\affiliation{\ICREA}

\author{Emilia Witkowska}
\affiliation{\IOPPAS}
 
\author{Jan Chwede\'nczuk}
\affiliation{\UW}

\begin{abstract}
  We demonstrate that the one-axis twisting (OAT), a versatile method of creating non-classical states of bosonic qubits,  is a powerful source of many-body Bell correlations. We develop a fully analytical and universal treatment of the process,  which allows us to identify the critical time at which the Bell correlations emerge, and predict the depth of Bell correlations at all subsequent times.
  Our findings are illustrated with a highly non-trivial example of the OAT dynamics generated using the Bose-Hubbard model. 
\end{abstract}

\maketitle
 
Non-classical correlations, namely entanglement and
Bell correlations are fundamental properties of the quantum many-body systems and crucial resources for emerging quantum technologies.  Because of enormous challenges in fault-tolerant quantum computing, the main goal for quantum technologies in the next decade is to generate, characterize, 
validate, and certificate massively correlated quantum
states \cite{Acin_2018,Eisert2020,Kinos2021,Laucht_2021,Zwiller2022,Fraxanet2022}.  In order to fully exploit many-body Bell correlations, we need an experimental protocol to generate such quantum states and a method for classifying the depth of many-body Bell correlations.
 

The well known method for generating entangled states is the one-axis twisting (OAT), which has been used in various configurations and is a subject of extensive theoretical studies~\cite{PhysRevA.46.R6797,PhysRevA.47.5138,PhysRevA.92.043622,PhysRevA.46.R6797, PhysRevLett.100.210401, Li2009, PhysRevA.96.013823, Kajtoch-sc-2018, Schulte2020, PhysRevLett.126.160402, PhysRevA.105.022625}. OAT can be realized with variety of ultra-cold systems, utilizing atom-atom collisions~\cite{Treutlein2010,Oberthaler2010,Chapman2012,PhysRevLett.125.033401}, and atom-light interactions~\cite{PhysRevLett.104.073602,PhysRevLett.105.080403}. Theoretical proposals for the OAT simulation with ultra-cold atoms in optical lattices~\cite{Kajtoch2018, PhysRevResearch.1.033075, Plodzien2020, PhysRevResearch.3.013178,Hernandez2022}, which  effectively simulate Hubbard and Heisenberg models, are awaiting for proof-of-principle experimental demonstration. 
 
It is well understood that OAT creates many-body entangled states 
and the two-body Bell correlations  inherent in two-body correlations ~\cite{Tura1256,schmied2016bell}. 
An important question in OAT procedure is about generation of the many-body Bell correlated states. In \cite{Aloy2019, Baccari2019, Tura2019, PRXQuantum.2.030329} authors provide set of Bell inequalities based on second-order correlators, and show that their violation implies $k$-producibility of non-locality with $k\leqslant6$ for large number of parties. 

Here we address the problem of nonlocality for an arbitrary depth $k$ in the collection of qubits subject to the OAT procedure by employing a wide family 
of Bell inequalities using many-body correlators ~\cite{zukowski2002bell,cavalcanti2007bell,he2011entanglement,cavalcanti2011unified,spiny.milosz,PhysRevLett.126.210506,10.21468/SciPostPhysCore.5.2.025}. 
We analytically evaluate the many-body correlator 
providing a powerful formula allowing to characterize the depth of many-body nonlocality at any moment of time. As such, we indicate the critical time at which the many-body Bell correlations emerge.

We begin by a brief outline of the OAT dynamics of $N$ bosons. Each particle has two internal states $a$ and $b$, with bosonic operators $\hat a$ and $\hat b$ anihilating a particle in a given state.
The system is conviniently described by means of the collective angluar momentum (spin) of length $j=N/2$, and the corresponding operators
$\hat J_x=\frac12\left(\hat a^\dagger\hat b+\hat a\hat b^\dagger\right)$, 
$\hat J_y=\frac1{2i}\left(\hat a^\dagger\hat b-\hat a\hat b^\dagger\right)$,
$\hat J_z=\frac12\left(\hat a^\dagger\hat a-\hat b^\dagger\hat b\right)$.
The OAT Hamiltonian reads
\begin{align}\label{eq.ham}
  \hat H_{\rm oat}=\chi\hat J_z^2,
\end{align}
where $\chi$ is an energy-unit constant~\cite{PhysRevA.47.5138}. 
The implementation of the OAT begins with a spin coherent state (CSS), 
$\ket{\frac\pi2,\varphi}_{\rm css}=\frac1{\sqrt{N!}}\left[\frac{\hat a^\dagger +e^{i\varphi}\hat b^\dagger}{\sqrt{2}}\right]^N\ket0$ being an eigen-state of 
$\hat J_x$ for $\varphi = 0$, $\hat J_x\ket{\frac\pi2,0}_{\rm css}=\frac N2\ket{\frac\pi2,0}_{\rm css}$.
This state undergoes the dynamics 
\begin{align}\label{eq.state}
  \hat\varrho(\tau)=\sum_{n,m=-\frac N2}^{\frac N2}
  c_n c_m
  e^{-i(n^2-m^2)\tau}\ketbra{n}{m},
\end{align}
with the period equal to $\pi$ for dimensionless $\tau=t\chi/\hbar$~for even $N$. Here, $\ket{n}$ is an eigen-state of $\hat J_z$, namely $\hat J_z\ket n=n\ket n$ with $\frac N2-n$ bosons in mode $a$ and $\frac N2+n$ in $b$, while $c_n = 2^{-N/2}\sqrt{\binom{N}{\frac{N}{2} + n}}$. 
%
For times $\tau \lesssim \tau_{\rm s}$, $\tau_{\rm s}\approx  N^{-2/3}$~\cite{PhysRevA.47.5138,https://doi.org/10.48550/arxiv.2112.01786}
the spin squeezing is generated, quantified by the
squeezing parameter
\begin{align}\label{eq.spin}
  \xi^2=N\frac{\Delta^2 \hat J_{\perp, {\rm min}}}{\av{\hat J}^2},
\end{align}
here $\av{\hat J}$ is the length of the mean collective spin and $\Delta^2\hat J_{\perp,{\rm min}}\equiv\av{\hat J_{\perp,{\rm min}}^2}-\av{\hat J_{\perp,{\rm min}}}^2$ 
is the minimal variance of the collective spin orthogonally 
to its direction~\cite{wineland1994squeezed}.
When $\xi^2$ drops below unity, it signalls the presence of entanglement between the qubits and the potential metrological gain with respect to the standard quantum limit.

At later times $\tau>\tau_{\rm s}$ the state enters a non-Gaussian regime and at $\tau_q=\pi/q$  for $q=2,4,6\ldots$, a macroscopic superposition of coherent states~\cite{2008Ferrini,2014Pawlowski}
\begin{align}\label{eq.state.q}
  \ket{\psi_q}=\frac1{\sqrt q}\sum_{k=0}^{q-1}e^{i\tau_qk(k+N)}\ket{\frac\pi2,2\tau_qk}_{\rm css}
\end{align}
is created. In particular, for $\tau_q = \pi/2$ a superposition of two coherent state is realized, forming the  Schr\"odinger cat  state.

To quantify the extent of many-body Bell correlations generated in the OAT process, we use a quantum $N$-body
correlator $\tilde{\mathcal E}_N^{(q)}$ which witnesses the Bell correlations if the inequality
\begin{align}\label{eq.bell.bos}
  \tilde{\mathcal E}^{(q)}_N=\modsq{\frac1{N!}\av{\hat J_+^N}}\leqslant 2^{-N}.
\end{align}
is violated~\cite{10.21468/SciPostPhysCore.5.2.025}. Here $\hat J_+=\hat J_1+i\hat J_2$, where $1$ and $2$ denote any two orthogonal directions spanned by the triple operators $\{ \hat{J}_{x},\hat{J}_{y},\hat{J}_{z} \}$. 
For the derivation of this bound, see the Appendix~\ref{app.bell} and References~\cite{cavalcanti2007bell,he2011entanglement,cavalcanti2011unified,10.21468/SciPostPhysCore.5.2.025}. 
The average
$\av{\hat J_+^N}$ determines the coherence between the extreme elements of the density matrix, namely between $\ket{-\frac N2}$ and $\ket{\frac N2}$, 
where these two kets are the eigenstates of the operator orthogonal to directions $1$ and $2$. 

A natural choice of the orientation of the plane spanned by the $\hat J_1$ and $\hat J_2$ operators is such that maximizes the Bell correlator~$\tilde{\mathcal E}^{(q)}_N$. This, in turn, is determined
by the OAT Hamiltonian~\eqref{eq.ham} and the state~\eqref{eq.state}. By inspecting Eq.~\eqref{eq.state.q}, we notice that at times $\tau_q$, the OAT generates superpositions of
eigenstates of $\hat J_x$ with varying weigths and eigenvalues, depending on $q$. The ultimate and most quantum product of OAT is at $\tau_2=\frac\pi2$, i.e., 
\begin{align}
  \ket{\psi_2}=\frac1{\sqrt 2}\left(\ketL{\frac N2}_x+\ketL{-\frac N2}_x\right),
\end{align}
hence a NOON state. Here, the subscript $x$ denotes specifically that the two components of the superposition are the eigen-states of $\hat J_x$.
This observation, namely that the OAT procedure creates superpositions of eigenstates of $\hat J_x$, indicates that to detect most of quantum features of the OAT state, 
one should align $1$ and $2$ in the plane orthogonal to $x$. However, it is more convenient (and equivalent), to align $1$ and $2$ in the $x$-$y$ plane and
rotate the OAT state~\eqref{eq.state} around the $y$ axis through angle $\frac\pi2$, 
\begin{align}
  \hat\varrho_{\rm rot}(\tau)=e^{-i\frac\pi2\hat J_y}\hat\varrho(\tau)\,e^{-i\frac\pi2\hat J_y}=\sum_{n,m=-\frac N2}^{\frac N2}\tilde\varrho_{nm}^{(\tau)}\ketbra{n}{m},
\end{align}
where the transformed element of the density matrix is
\begin{align}
  \tilde\varrho_{nm}^{(\tau)}=\sum_{n',m'=-\frac N2}^{\frac N2}d_{nn'}^{\frac N2}\left(\frac\pi2\right)d_{mm'}^j\left(\frac\pi2\right)\varrho_{n'm'}^{(\tau)}.
\end{align}
Here, $d_{\alpha\beta}^j(\phi)$ denotes the element of the Wigner rotation matrix
~\footnote{The matrix element of the Wigner matrix is $d_{nm}^j(\theta)=\bra n e^{-i\theta\hat J_y}\ket m=\sqrt{\frac{(j+m)!(j-m)!}{(j+n)!(j-n)!}}\left(\sin\frac\theta2\right)^{m-n}\left(\cos\frac\theta2\right)^{m+n}$,
with $j=\frac N2$.}.

After the rotation, the eigenstates of $\hat J_x$ transform into the eigenstates of $\hat J_z$ and the proper choice of the rising operator
to maximize $\tilde{\mathcal E}^{(q)}_N$ is $\hat J_+=\hat J_x+i\hat J_y$, which gives
\begin{align}\label{eq.coeff.bell}
  \tilde{\mathcal E}^{(q)}_N=\modsq{\tilde\varrho_{-\frac N2,\frac N2}^{(\tau)}}.
\end{align}
For the state (\ref{eq.state}) we obtain $\tilde\varrho_{-\frac N2,\frac N2}^{(\tau)}=\tilde C_{-\frac N2}^{(\tau)}\tilde C_{\frac N2}^{(\tau)}$, 
where
\begin{subequations}\label{eq.tildas}
  \begin{align}
    &\tilde C^{(\tau)}_{-\frac{N}{2}}=\frac{1}{2^{N}}\sum_{n=-\frac{N}{2}}^\frac{N}{2}{N\choose {n+\frac N2}}e^{-i\tau n^2},\\
    &\tilde C^{(\tau)}_{\frac{N}{2}}=\frac1{2^{N}}\sum_{n=-\frac{N}{2}}^{\frac{N}{2}}{N\choose {n+\frac{N}{2}}}(-1)^ne^{-i\tau n^2}
  \end{align}
\end{subequations}
are the two coefficients of the state expressed in the basis of eigenstates of $\hat J_z$ with two extreme eigenvalues. 

First, we focus on the short-time dynamics to identify the critical instant $\tau_{\rm crit}$ at which the Bell correlations emerge in the OAT procedure. To this end, we notice that when
$\tau$ is short and $N$ is large, the sums in Eqs~\eqref{eq.tildas} can be evaluated by approximating the binomial function with a Gaussian, namely
\begin{align}
  \frac1{2^N}\binom{N}{n+\frac N2}\simeq\sqrt{\frac2{\pi N}} e^{-\frac2Nn^2},
\end{align}
which gives
\begin{align}
  \tilde C^\alpha_{-\frac N2}\simeq\frac1{\sqrt{1+i\kappa}},\ \ \ \tilde C^\alpha_{\frac N2}\simeq2\frac{e^{N\pi\frac{i(\kappa+\frac\pi4)+1}{2(\kappa-i)}}}{\sqrt{1+i\kappa}}
\end{align}
and the Bell correlator becomes
\begin{align}\label{eq:short-times}
  \tilde{\mathcal E}^{(q)}_N\simeq\frac4{(1+\kappa^2)^2}e^{-\frac{\pi^2N}{8(1+\kappa^2)}}
\end{align}
with $\kappa=\frac{\tau N}2$. The $\tau_{\rm crit}$ can be obtained by comparing the logarithm of $\mathcal E_N$ 
with the logarithm of the threshold attainable by local realistic theories [see Eq.~\eqref{eq.bell.bos}], namely
\begin{align}
  \ln4-2\ln(1+\kappa^2)-\frac{\pi^2N}{8(1+\kappa^2)}=-N\ln2.
\end{align}
Note that for large $N$, the two logarithms on the left-hand-side can be neglected giving the critical time at which the Bell correlations are created in the OAT procedure
\footnote{A variation of the Bell correlator (\ref{eq:short-times}) in time exhibits a single maximum. We are interested here in the grow of $\tilde{\mathcal E}^{(q)}_N$ for short times which is determined by the exponent in (\ref{eq:short-times}). The very short time variation of Bell correlator can be well approximated by $4e^{-\frac{\pi^2N}{8(1+\kappa^2)}}$ what gives (\ref{eq:tcrit}).}
\begin{align}\label{eq:tcrit}
  \tau_{\rm crit} \simeq \frac2{N}\sqrt{\frac{\pi^2}{8\ln2}-1}\approx\frac{1.77}N.
\end{align}
It is worth to stress here, that the Bell correlations emerge before optimal squeezing time as $\tau_s < \tau_{\rm crit}$ in the large $N$ limit. 
At the instant $\tau_{\rm crit}$ three particles are Bell correlated, the smallest minimal number for which the correlator~\eqref{eq.bell.bos} exceeds the local bound~\cite{spiny.milosz}.

At later times, the continous approximation fails, as it cannot capture the genuinely quantum discrete-$N$ interference phenomena.  
A close analogy is the series of collapses and revivals in the two-level dynamics of
an atom driven by a quantized coherent electromagnetic fiels~\cite{mandel1995optical}, which are not captured by the semi-classical approach.

However, this is when the result of Eq.~\eqref{eq.state.q} comes at hand. At instants $\tau_q$ when superpositions of coherent states (\ref{eq.state.q}) are formed, 
the two extreme coefficients can be calculated analytically, giving
\begin{subequations}\label{eq.cs}
  \begin{align}
    &\tilde C_{-\frac N2}^{(\tau_q)}=\frac1{\sqrt q}\sum_{l=0}^{q-1}e^{i\tau_ql^2}\cos^N(\tau_ql),\\
    &\tilde C_{\frac N2}^{(\tau_q)}=\frac{i^N}{\sqrt q}\sum_{l=0}^{q-1}e^{i\tau_ql^2}\sin^N(\tau_ql).
  \end{align}
\end{subequations}
Note that the sine and cosine functions in Eqs.~\eqref{eq.cs}, when taken to the power of $N$, give non-zero values only when they are close to unity,
which is for $l=q/2$ and $l=0$, respectively. This gives the Bell coefficient as
\begin{align}\label{eq.approx}
  \tilde{\mathcal E}^{(q)}_N\simeq\frac1{q^2}
\end{align}
for times $\tau_q$. This is a very simple yet powerful formula allowing to predict the extent of Bell correlations in various many-body systems.

Hence, as $q$ drops, so that the time grows (recall that $\tau_q=\frac\pi q$), the value of the Bell correlator increases to reach the maximal attainable value $\tilde{\mathcal E}^{(q)}_N=\frac14$
at the half of the dynamics period. But there is more information about the many-body Bell correlations that can be extracted from the expression~\eqref{eq.approx}. Namely, when
\begin{align}\label{eq.hier}
  \tilde{\mathcal E}^{(q)}_N>\frac18\frac1{2^{N-k}},
\end{align}
the correlator can be reproduced with a system of $N$ qubits, where the Bell correlations encompass at least $k$ qubits (in analogy to $k$-partite entanglement)~\cite{spiny.milosz,PhysRevLett.126.210506}.
For instance, when 
$\tilde{\mathcal E}^{(q)}_N>\frac18$, all $N$ qubits are Bell-correlated, if $\tilde{\mathcal E}^{(q)}_N>\frac1{16}$, the Bell correlations extend over at least $N-1$ particles and so on.
Note that the correlator from Eq.~\eqref{eq.bell.bos} can also be used to witness the $k$-partite entanglement in the system~\cite{cavalcanti2011unified,spiny.milosz}. 
Similarly to Eq.~\eqref{eq.bell.bos}, when
\begin{align}\label{eq:entangl}
  \tilde{\mathcal E}^{(q)}_N\leqslant4^{-N},
\end{align}
the many-body correlator can be reproduced with a fully-separable state. The analogy to the $k$-partite Bell correlations extends further, namely if
\begin{align}\label{eq:kpartite}
  \tilde{\mathcal E}^{(q)}_N>\frac1{16}\frac1{4^{N-k}},
\end{align}
the correlator is consistent with that of a system where $k$ qubits form a $k$-partite entangled state and the other $N-k$ are separable, just as in Eq.~\eqref{eq.hier}.

Since the expression~\eqref{eq.state.q}, which is used to derive~\eqref{eq.approx} is valid for $\tau_q> \tau_s$, hence we observe that at the shortest time when Eq.~\eqref{eq.approx}
can be used, it holds that $\tilde{\mathcal E}^{(q)}_N\simeq N^{-\frac43}$.
For instance, when $N=10^3$, this gives
$\tilde{\mathcal E}^{(q)}_N\simeq10^{-4}$.
Using Eq.~\eqref{eq.hier}, we obtain that in this case $\tilde{\mathcal E}^{(q)}_N\simeq\frac1{10^4}>\frac18\frac1{2^{11}}$. Hence even at such short time, the Bell correlations
extend over $k = 10^3-11=989$ particles. 
Thus we show that the OAT procedure naturally generates many-body Bell correlated states,
very early in dynamics when $\tau \gtrsim \tau_{\rm crit}$, one of the main results of this work.

\begin{figure}[]
{\includegraphics[width=\linewidth]{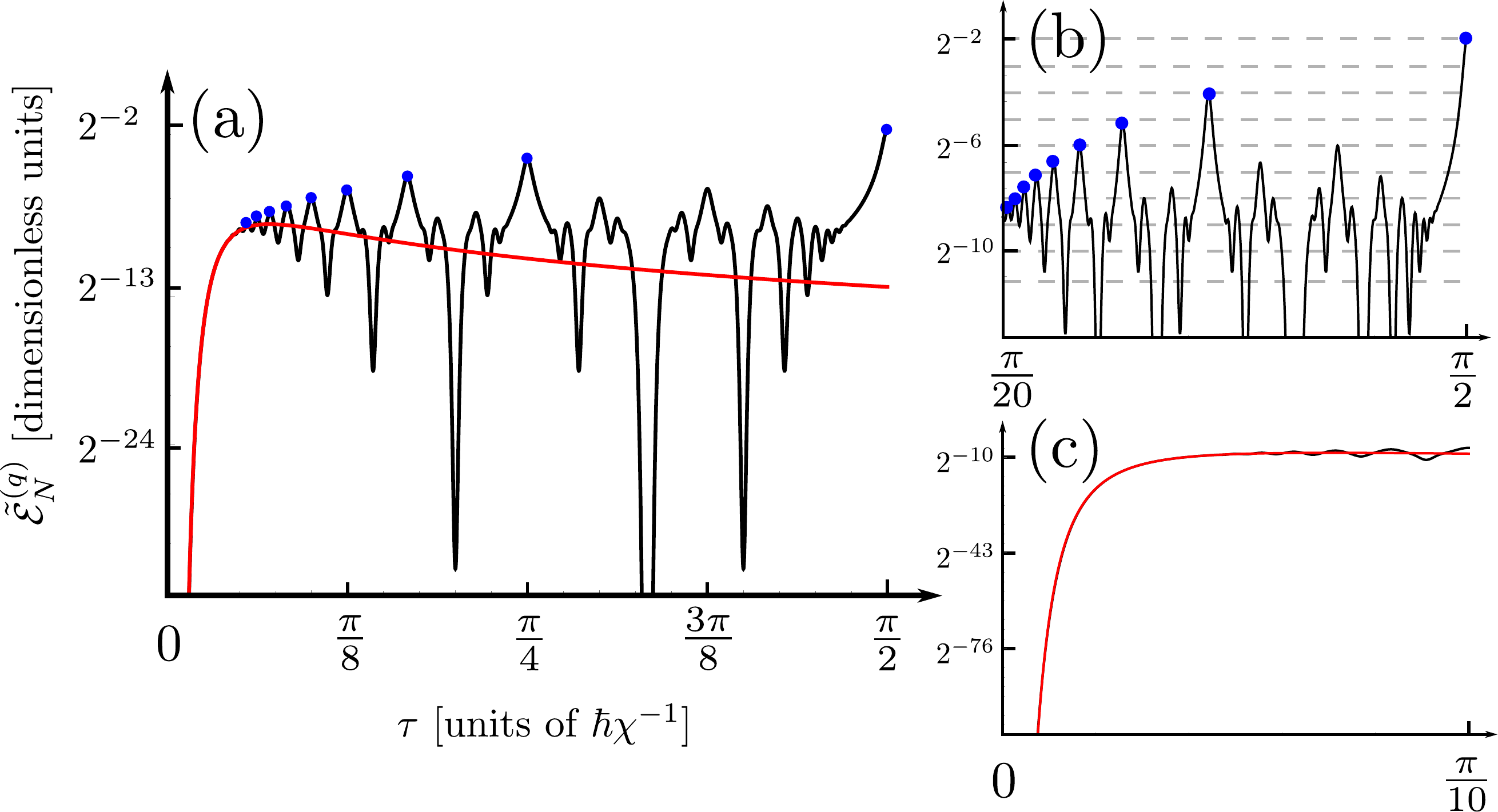}}
\caption{The Bell correlator $\tilde{\mathcal{E}}_N$ (black solid line) as a function of $\chi t$ and for $N=200$. (a): Compared with the short-times approximate behaviour~\eqref{eq:short-times} (solid red line) 
  and the long-times solution from Eq.~\eqref{eq.approx} (blue points). 
  (b) The zoom onto the long-time behaviour and the growing  depth of Bell correlations signalled with dashed grey lines for $k=N, N-1, \dots, N-8$ from top to bottom. 
  (c) The focus onto short times up to $\chi t/\hbar =\pi/10$.
}
\label{fig:fig1}
\end{figure}

Next, we illustrate our theory for certification of Bell correlations with OAT in the specific system, namely $N$ ultra-cold bosonic atoms in 1D optical lattice with $M=N$ sites, each of mass $m$ in two internal states $a$ and $b$. 
The system Hamiltonian reads
\begin{align}\label{H_BH}
  &\hat H=\hat{H}_{a} + \hat{H}_b + \hat{H}_{ab},
\end{align}
where
\begin{align}
  \hat H_a=-J\sum_{j} (\hat a_j^\dagger\hat a_{j+1}+\mathrm{h.c.}) + \frac{U_{aa}}{2}\sum_j\hat{n}_{a,j}(\hat{n}_{a,j}-1)
\end{align}
describes nearest-neighbor tunneling and repulsive interaction as for the Bose-Hubbard model, and analogically for $\hat H_b$, while $\hat H_{ab}=U_{ab}\sum_{j}\hat{n}_{a,j}\hat{n}_{b,j}$.  Here $\hat a_j$ is the annihilation operator of the atom at the $j$th lattice site in mode $a$
and $\hat{n}_{a,j}$ is the corresponding particle number operator. We assume the optical lattice is formed by standing laser beam with 
wave-vector $k_{\rm latt} = 2\pi/\lambda$, where the optical lattice wavelength is $\lambda = 2d$, $d$ is distance between neighboring sites. The height of  the optical lattice $V_0$ 
determines the coupling parameters, \textit{i.e.} hopping amplitude
$J  = 4/\sqrt{\pi} V_0^{3/4} e^{-2\sqrt{V_0}}$, and contact interaction amplitudes $U_{\alpha\alpha'}  = \sqrt{8/\pi} a_{\alpha\alpha'}V_0^{1/4}$, with $a_{\alpha\alpha'}$ being the scattering lengths.  
We consider the parameters used in~\cite{Plodzien2020} which can be realized in current experiments, namely symmetric intra-component interactions, $U_{aa} = U_{bb} \equiv U$, and inter-component interaction to be $U_{ab}=0.95U$. 
The recoil energy $E_R = \hbar^2k^2/2m$ is set as an energy unit.
The collective spin operators can be defined through 
$\hat{J}_+ = \sum_{j=1}^{M}\hat{a}_{j}^\dagger\hat{b}_j$, $\hat{J}_- = \sum_{j=1}^{M}\hat{b}_{j}^\dagger\hat{a}_j$,
and $\hat{J}_x = (\hat{J}_+ + \hat{J}_-)/2$, 
$\hat{J}_y = (\hat{J}_+ - \hat{J}_-)/(2 i)$,  
$\hat{J}_z =(\hat{N}_{a} - \hat{N}_{b})/2$ where $\hat{N}_{a/b}$ is the operator of number of atoms in the state $a/b$. 

The OAT model can be simulated with the BH model (\ref{H_BH}) in the superfluid phase when the condensate fraction is close to one \cite{Kajtoch2018,Plodzien2020}. To see this, we consider (\ref{H_BH}) in the Fourier space, 
by using $\hat{a}_{j}= \frac1{\sqrt{M}} \sum_{q_n} e^{- i x_j q_{n}} \hat{a}_{q_n}$ with $q_n = \frac{2 \pi}{N}n$ and $n=0,\pm1, \pm2, \cdots$ is an integer (and analogically for $\hat{b}_j$).
Next, when atoms microscopically occupy the zero momentum mode $q_n=0$, the system Hamiltonian considered for zero quasi-momentum mode $q_n=0$ reduces to the OAT model,
\begin{equation}\label{eq.oat.app}
  \hat{H} \approx \chi \hat{J}_{z,q_n=0}^2,
\end{equation}
when omitting constant energy terms, and where $\chi = \frac{U - U_{ab}}{M}$ gives the relevant time-scale.

\begin{figure}[]
  \centering
  \includegraphics[width=\linewidth]{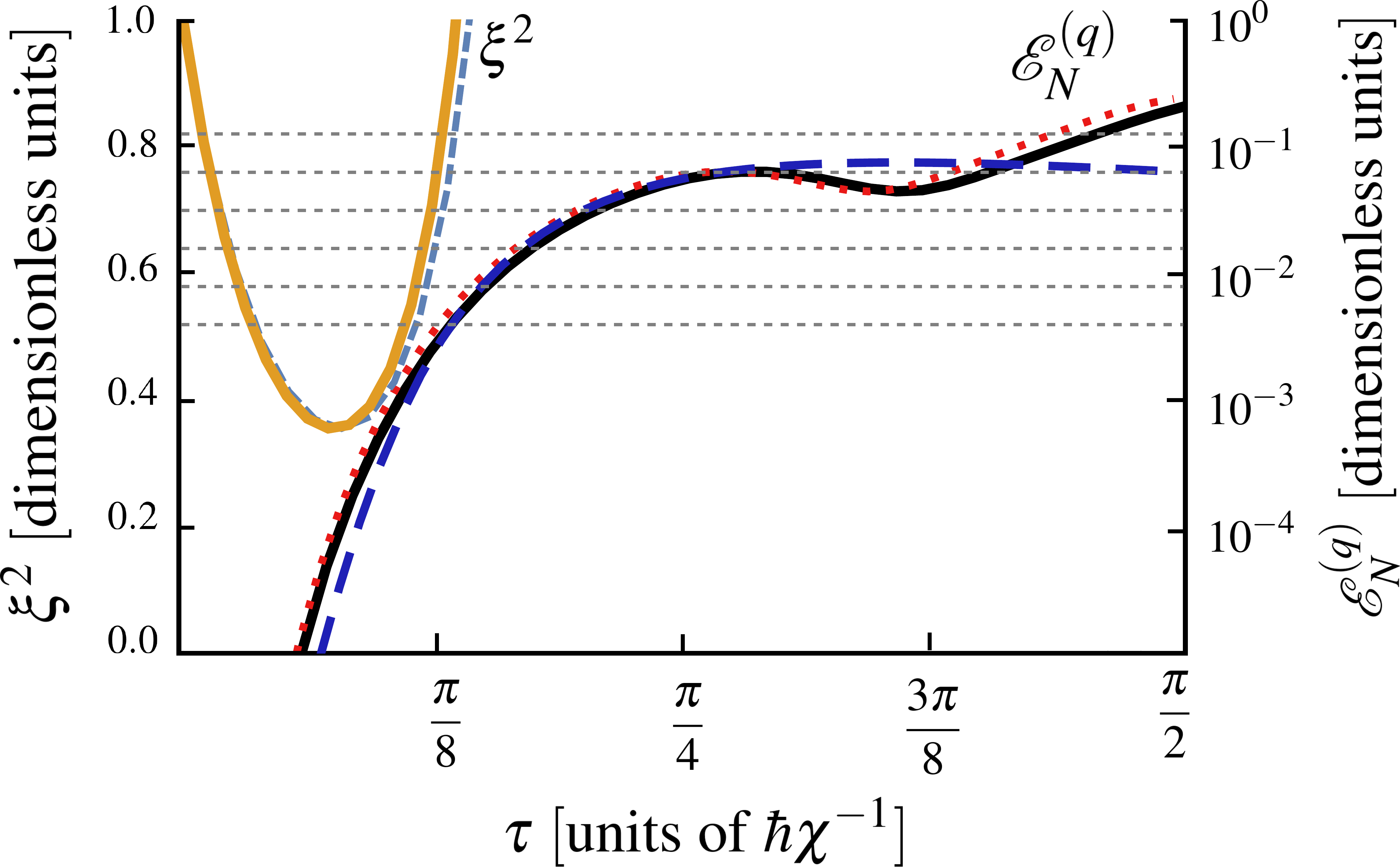}
  \caption{The Bell correlator $\tilde{\mathcal E}_N^{(q)}$ (the three curves and the vertical scale in the right part of the plot)
  	and the spin-squeezing parameter $\xi^2$ (the two curves and the vertical scale on the left) for $N=8$. For $\tilde{\mathcal E}_N^{(q)}$: 
  	the solid black line is the correlator from Eq.~\eqref{eq.coeff.bell}, for the two component BHM \eqref{H_BH} when the system is in the superfluid phase. 
  	The dotted red curve is obtained for the OAT model, see Eq.~\eqref{eq.oat.app}, while the dashed blue curve shows the result of the short-times approximation  (~\ref{eq:short-times}).
  	Particular bounds for the $k$-qubit Bell correlations~\eqref{eq.hier} are marked by the dashed grey lines for $k=8, 7, \dots, 3$ from top to bottom. For $\xi^2$: the squeezing parameter~\eqref{eq.spin} is marked by the solid orange line (exact numerical results for BHM), while the dashed blue line is calculated with the OAT\eqref{eq.oat.app}.}
  \label{fig:fig2}
\end{figure}

We performed many-body numerical calculations \footnote{We employ a standard density matrix renormalization group (DMRG) technique for calculating initial spin coherent state \cite{PhysRevLett.69.2863,PhysRevB.48.10345,RevModPhys.77.259,SCHOLLWOCK201196,ORUS2014117}. Time evolution was prepared within algorithm for time evolution where one-site time-dependent variational principle (TDVP) scheme \cite{Kramer_2008, PhysRevLett.107.070601, PhysRevLett.109.267203,PhysRevB.94.165116} is combined with a global basis expansion \cite{PhysRevB.102.094315}. To perform both DMRG and time evolution we use ITensor C++ library \cite{fishman2021itensor}, where we employ the codes for modified TDVP provided by the authors
of \cite{PhysRevB.102.094315}. 
}, to prepare the initial spin coherent state given by the symmetric superposition of atoms in states $a$ and $b$,  
to evaluate the unitary evolution and calculate the  
spin squeezing parameter (\ref{eq.spin}) and the Bell correlator (\ref{eq.bell.bos}). 
In Fig.~\ref{fig:fig2} we show results for $N=8$ atoms. One can clearly see the overall agreement between the Bose-Hubbard  (solid lines) and OAT (closed points) models. 
However, an analysis of the evolution of $\tilde{\mathcal{E}}^{(q)}_N$  brings interesting conclusions. 
The lower bound of $\tilde{\mathcal{E}}^{(q)}_{N=8} \approx 4\times  10^{-3}$ is marked in Fig.\ref{fig:fig2}, and Bell correlations emerge for times $\tau_{\rm crit}\approx \pi/8$ according to (\ref{eq:tcrit}).
We can see that around this time, the states are spin squeezed as $\xi^2<1$ (marked by the orange line) what suggests that there are already non-trivial two-body correlations in the system.

In the subsequent moments of time Bell correlations start to extend over at least three atoms, and next four atoms when $\tilde{\mathcal{E}}^{(q)}_{N=8} \gtrsim 7.8\times 10^{-3}$, five atoms when 
$\tilde{\mathcal{E}}^{(q)}_{N=8} \gtrsim 1.5\times 10^{-2}$, etc., and over all eight atoms when $\tilde{\mathcal{E}}_{N=8} \gtrsim 0.125$.
Finally, let us comment about the depth of entanglement given by (\ref{eq:entangl}) 
 i.e., $\tilde{\mathcal{E}}^{(q)}_{N=8} \gtrsim 1.5 \times 10^{-5}$, which is surpassed for $\tau\approx\pi/20$. 
According to Eq.~\eqref{eq:kpartite}, when $\tilde{\mathcal{E}}^{(q)}_{N=8} \gtrsim 2\times 10^4$ the entanglement extends over four atoms. Finally 
when $\tilde{\mathcal{E}}^{(q)}_{N=8}>\frac1{16}$ entanglement extends over the whole system.

In this work  we presented a systematic analytical study of the creation of many-body Bell-correlated states generated during one-axis twisting dynamics in two-component bosonic systems. 
We identified the critical time at which the many-body Bell correlations emerge and derived a simple and powerful formula allowing to characterize the Bell correlations- and entanglement-depth at later times. 
We applied these findings to clasify the generation of many-body Bell correlations in systems of two-component bosons loaded into a one-dimensional optical lattice. 
We showed that our analytical findings are in a very good agreement with the full many-body numerical calculations. 

The experimental verification of our findings requires access to many-body quantum correlation functions. A remarkable progress in experimental advances in control of many-body quantum systems  allows measurement of correlation functions up to $6$-th order \cite{Dall2013}, $2$-nd  R\'enyi entropy  for $N=4$ \cite{Islam2015}, and $N=5$ particles \cite{Linke2018} 
via extraction  characteristic of a quantum state using a controlled-swap gate acting on two copies of the state \cite{Eckert2002}, and for $N=10$ particles \cite{Brydges2019} via randomized measurements technique \cite{Vermersch2018,Elben2019,Elben2020,Elben2020_PRL,Rath2021,Elben2022}. However, for the indistinguishable particles considered here, the measurement of the proposed many-body correlator still presents an experimental challenge for $ N\ge 6$ atoms. The future research direction would be to consider single addressable particles, like trapped ions, and consider generation and  detection of many-body Bell correlations in NISQ devices.

Our study contributes to the dynamically emerging field of quantum technologies having both fundamental and practical aspects.

\section*{ACKNOWLEDGMENTS}
This work was supported by QuantEra project MAQS, Grant No. UMO-2019/32/Z/ST2/00016 (E.W.).
J.Ch. was funded by the National Science Centre, Poland, within the QuantERA II Programme that has received funding from the European Union’s Horizon 2020 research and innovation programme under Grant Agreement No 101017733, Project No. 2021/03/Y/ST2/00195.
M.P. acknowledges the support of the Polish National Agency for Academic Exchange, the Bekker programme no: PPN/BEK/2020/1/00317, and the computer resources at MareNostrum and the technical support provided by BSC (RES-FI-2022-1-0042).
ICFO group acknowledges support from: ERC AdG NOQIA; Agencia Estatal de Investigación (R\&D project CEX2019-000910-S, funded by MCIN/ AEI/10.13039/501100011033, Plan National FIDEUA PID2019-106901GB-I00, FPI, QUANTERA DYNAMITE PCI2022-132919, Proyectos de I+D+I “Retos Colaboración” QUSPIN RTC2019-007196-7), MCIN via European Union NextGenerationEU (PRTR); Fundació Cellex; Fundació Mir-Puig; Generalitat de Catalunya through the European Social Fund FEDER and CERCA program (AGAUR Grant No. 2017 SGR 134, QuantumCAT \ U16-011424, co-funded by ERDF Operational Program of Catalonia 2014-2020); EU Horizon 2020 FET-OPEN OPTOlogic (Grant No 899794); National Science Centre, Poland (Symfonia Grant No. 2016/20/W/ST4/00314); European Union’s Horizon 2020 research and innovation programme under the Marie-Skłodowska-Curie grant agreement No 101029393 (STREDCH) and No 847648 (“La Caixa” Junior Leaders fellowships ID100010434: LCF/BQ/PI19/11690013, LCF/BQ/PI20/11760031, LCF/BQ/PR20/11770012, LCF/BQ/PR21/11840013).
\appendix
\section{Many-body Bell inequality}\label{app.bell}
To quantify the strength  of many-body Bell correlations generated in the OAT process, we use a broad family of Bell inequalities first introduced in 
Refs~\cite{cavalcanti2007bell,he2011entanglement,cavalcanti2011unified}. When each of $N$ parites measures two binary quantities $\sigma_{1,2}^{(k)}=\pm1$ (with $k=1\ldots N$),
then the correlator
\begin{align}
  \mathcal E_N=\modsq{\av{\sigma_+^{(1)}\ldots\sigma_+^{(N)}}},
\end{align}
with $\sigma_+^{(k)}=\frac12(\sigma_1^{(k)}+i\sigma_2^{(k)})$ can be reproduced by a theory consistent with the postulates of local realism if it takes the form
\begin{align}
  \mathcal E_N=\modsq{\int\!\! d\lambda\, p(\lambda)\sigma_+^{(1)}(\lambda)\ldots\sigma_+^{(N)}(\lambda)},
\end{align}
where $\lambda$ is a hidden variable and $p(\lambda)$ is its probability distribution. Using the Cauchy-Schwarz inequality we obtain
\begin{align}\label{eq.bell.ineq}
  \mathcal E_N\leqslant\int\!\! d\lambda\, p(\lambda)\modsq{\sigma_+^{(1)}(\lambda)\ldots\sigma_+^{(N)}(\lambda)}=\frac1{2^N},
\end{align}
which is the $N$-body Bell inequality. 

For quantum systems, $\sigma_{1,2}^{(k)}$ are replaced with the Pauli operators $\hat\sigma_{1,2}^{(k)}$ and $\mathcal E_N$ is replaced by $\mathcal E^{(q)}_N$, which is a quantum correlator, i.e., 
\begin{align}
  \mathcal E^{(q)}_N=\modsq{\av{\hat\sigma_+^{(1)}\ldots\hat\sigma_+^{(N)}}},
\end{align}
where $\hat\sigma_+^{(k)}$ is a rising operator. If $\mathcal E^{(q)}_N$ violates the bound from Eq.~\eqref{eq.bell.ineq}, it witnesses the many-body Bell correlations. 

For bosonic systems, qubits cannot be addressed individually, thus the $\hat\sigma_+^{(k)}$'s must be replaced with 
collective angular momentum operators. Formally, this is achieved by symmetrizing the product of $N$ operators $\hat\sigma_+^{(1)}\ldots\hat\sigma_+^{(N)}$. Since
all orderings of these operators are equivalent and there are $N!$ such settings, the Bell inequality~\eqref{eq.bell.ineq} for bosonic systems takes the form
\begin{align}
  \tilde{\mathcal E}^{(q)}_N=\modsq{\frac1{N!}\av{\hat J_+^N}}\leqslant 2^{-N}.
\end{align}
A more detailed discussion of this derivaiton can be found in~\cite{10.21468/SciPostPhysCore.5.2.025}. 

\bibliography{bibl}

\end{document}